\documentclass[twocolumn,showpacs,aps,floatfix,prd]{revtex4}

\usepackage{graphicx}
\usepackage{dcolumn}
\usepackage{amsmath}
\usepackage{epsfig}
\usepackage{hhline}
\usepackage{amssymb}

\long\def\inst#1{\par\nobreak\kern 4pt\nobreak
    {\it #1}\par\vskip 10pt plus 3pt minus 3pt}

\begin{document}



\title{\large \bf
\Large \bf Constraining the Lattice Fluid Dark Energy from
SNe Ia, BAO and OHD}

\author{Xiaoxian Duan}\email{duan@bao.ac.cn}
\author{Yichao Li}\email{ycli@bao.ac.cn}
\author{Changjun Gao}\email{gaocj@bao.ac.cn}
\affiliation{The National Astronomical Observatories, Chinese
Academy of Sciences, China}

\date{\today}

\begin{abstract}
\noindent

Sanchez and Lacombe have ever developed a lattice fluid theory
based on a well-defined statistical mechanical model. Taking the
lattice fluid as a candidate of dark energy, we investigate the
cosmic evolution of this fluid. Using the combined observational
data of Type Ia Supernova (SNe Ia), Baryon Acoustic Oscillations
(BAO) and Observational Hubble Data (OHD), we find the best fit
value of the parameter in the model, $A=-0.3_{-0.1}^{+0.1}$. Then
the cosmological implications of the model are presented.

\end{abstract}


\maketitle

\newpage
\newcommand{\secname}{}

\setcounter{footnote}{0}





\renewcommand{\secname}{Introduction}
\section{Introduction} 
\label{sec:\secname}

Increasing amount of cosmological observations have provided
convincing evidences that our Universe is undergoing a late time
acceleration. The observations of Type Ia Supernova (SNe Ia)
\cite{Perlmutter1999,Riess1998}, Cosmic Microwave Background (CMB)
\cite{Spergel2007,Komatsu2009,Komatsu2011}, the Baryon Acoustic
Oscillations (BAO) \cite{Eisenstein2005,Percival2007,Percival2009}
and the Hubble data based on differential ages of the galaxies
(OHD) \cite{Jimenez2002} definitely reveal the accelerating
expansion of the Universe. Then how to explain the acceleration of
the Universe? The simplest explanation is the cosmological
constant (CC) which is raised by Einstein in General Relativity.
The cosmological constant could fit the observations very well.
However, it suffers from two fundamental problems: the fine tuning
problem and the coincidence problem \cite{weinberg:2000}. The fine
tuning problem is the following. The cosmological constant is also
the vacuum energy in quantum field theory. Then the estimated
value of the vacuum energy $\rho\simeq \rho_p$ (where $\rho_p$ is
the Planck density) is greater than the observed value $\rho\simeq
10^{-123}\rho_p$ by $123$ orders of magnitude. The coincidence
problem is the following. The densities of the dark energy and
matter evolve differently as the Universe expands. But they are
comparable in the present-day Universe. It is a great coincidence
if there are no any internal connections between them.

Therefore, a lot of dynamical dark energy models have been
explored. They are mainly classified into several types: scalar
field theories, modified gravity theories, phenomenological
models, and fluid dark energy models. The first type includes
quintessence \cite{Wetterich1988,Ratra1988,Fujii1982,Carroll1998},
phantom \cite{Caldwell2002}, K-essence
\cite{Chiba2000,Armendariz-Picon2000,Armendariz-Picon2001},
tachyon \cite{Padmanabhan2002,Padmanabhan20022}, quintom
\cite{Feng05}. The second type includes $f(R)$ theory
\cite{Capozziello20022,Capozziello2003,Capozziello20032,Carroll2004,Nojiri2003},
DGP theory \cite{Dvali2000} and so on. The third type includes
holographic dark energy \cite{Li04}, the agegraphic dark energy
\cite{Wei07} and so on. Last but not the least, there are fluid
dark energy models mainly cover the chaplygin gas
\cite{Kamenshchik2001, Bento2002, Benaoum2002, Dev2003, Bilic2002,
Alcaniz2003, Chimento2004, Fabris2011, Malekjani2011} and the Van
der Waals gas \cite{Capozziello2002, Capozziello2005, Kremer2003}.

Kamenshchik {\em et al.} studied the standard Chaplygin gas
\cite{Kamenshchik2001} with the equation of state $p =
-\frac{A}{\rho}$ (A is a positive constant). The standard
Chaplygin gas could behave as a pressureless dust at early times
and as a cosmological constant at late times in the evolution of
the Universe. But it has difficulty in explaining the CMB
anisotropic observations. The situation could be alleviated in the
generalized Chaplygin gas \cite{Bento2002} which has the equation
of state $p = -\frac{A}{\rho^{\alpha}}$ with $\alpha$ severely
constrained, $0 \leqslant \alpha <0.2 $. In general, Benaoum
studied a new equation of state $p = A\rho - \frac{B}{\rho^{n}}$
($n \geqslant 1$, $A$ and $B$ are positive constants). It could
interpolate between standard fluids at high energy densities
 and Chaplygin gas fluids at low energy densities \cite{Benaoum2002}.
On the other hand, Ref.~\cite{Capozziello2002,Capozziello2005}
investigated the possibility of Van der Waals fluid as both the
dark matter and dark energy, and Ref.~\cite{Kremer2003} studied
the evolution of the Universe filled with a mixture of van der
Waals fluid and dark energy. The lattice fluid dark energy (LFDE)
investigated in this paper falls into this last type.

The equation of state for lattice fluid is derived by Sanchez and
Lacombe \cite{Sanchez1978}. One find that the lattice fluid model
could perfectly describe the thermodynamic properties of a wide
variety of fluids. On the other hand, the four laws of black hole
mechanics, which are analogous to those of thermodynamics, were
originally derived from the classical Einstein equation
\cite{bardeen:1973}. Until the discovery of the quantum Hawking
radiation \cite{hawking:1975}, one recognize that the analogy is
in fact an identity. So the nature of space, time and matter is
closely related to the thermodynamics. By turning the logic
around, we started from the thermoplastics in order to seek for
the candidate of dark energy. We find that the lattice fluid could
behave as a fluid with negative pressure which is exactly the
property of dark energy. Therefore we constrain the
LFDE from the observations of SNe Ia, BAO and
OHD.

The paper is organized into four sections. In section II, we
derive the equation for the cosmic evolution of LFDE
 in the background of Friedmann-Robertson-Walker (FRW)
Universe. In section III, using the observational data of SNe Ia,
BAO and OHD, we seek for the best fit value of the parameter in
the model. In section IV, we present the main results: the best
fit value of the dimensionless parameter $A$, the evolution of
deceleration parameter $q$, the statefinder parameters $r, \ s$
with the redshift $z$. Section V gives the conclusion and
discussion. We shall use the units in which the speed of light $c$
is set to unity.

\renewcommand{\secname}{Lattice Fluid Dark Energy}
\section{Lattice Fluid Dark Energy}
\label{sec:\secname}

The equation of state for LFDE can be written
as \cite{Sanchez1978}
\begin{equation}
p_{X}=-\frac{\rho_{X}^{2}}{\rho_{0}}-A\rho_{0}\left[\ln(1-\frac{\rho_{X}}{\rho_{0}})+\frac{\rho_{X}}{\rho_{0}}\right]\;,
\label{eq:eos}
\end{equation}
where $\rho_{X}$ and $p_{X}$ are the energy density and pressure
of the fluid, respectively. $A$ is a dimensionless constant and
$\rho_{0}$ is the present-day cosmic energy density defined by
$\rho_{0}=\frac{3H_{0}^{2}}{8\pi}$ with $H_{0}$ the present-day
Hubble parameter.

Now we study the evolution of the Universe filled with matter
(includes baryon matter and dark matter) and LFDE. 
The metric of FRW Universe is given by:
\begin{equation}
ds^{2}=-dt^{2}+a^{2}(t)\left[\frac{dr^{2}}{1-Kr^{2}}+r^{2}(d\theta^{2}+\sin^{2}\theta
d\phi^{2})\right]\;,
\end{equation}
where $a(t)$ is the scale factor and $K=+1,\ 0,\ -1$ describe the
topology of the Universe which correspond to closed, flat and open
Universe, respectively.

The scale factor evolves according to the Friedmann equation:
\begin{equation}
H^{2}=(\frac{\dot{a}}{a})^{2}=\frac{8\pi G}{3}\rho_{total}-\frac{K}{a^{2}}\;,
\label{eq:fri}
\end{equation}
where $\rho_{total}$ is the total energy density of the Universe.
Observations reveal that the Universe is highly flat in space. So
we put $K=0$ in the following. For the matter-plus-dark energy
dominated Universe, we can safely neglect the contribution of
relativistic matter. Then the Friedmann equation can be written as
\begin{equation}
3H^{2}=8\pi G\left(\frac{\rho_{m}}{a^{3}}+\rho_{X}\right)\;,
\label{eq:flatfri1}
\end{equation}
where $\frac{\rho_{m}}{a^3}$ is the energy density of matter and
$\rho_{m}$ is the energy density of matter in the present-day
Universe.

In order to obtain $\rho_{X}$, we substitute the equation of state
for LFDE Eq.~(\ref{eq:eos}) into the energy conservation
equation:
\begin{equation}
\frac{d\rho_{X}}{da}+\frac{3}{a}\left(\rho_{X}+p_{X}\right)=0\;.
\label{eq:energycon1}
\end{equation}

Keeping in mind the relation $\frac{1}{a}=1+z$ and defining
$\Omega_{X}\equiv\frac{\rho_{X }}{\rho_{0}}$, we obtain the energy
conservation equation as follows
\begin{equation}
\frac{d\Omega_{X}}{dz}-\frac{3}{1+z}\left\{\Omega_{X}-\Omega_{X}^{2}-A\left[\ln\left(1-\Omega_{X}\right)+\Omega_{X}\right]\right\}=0\;.
\label{eq:energycon2}
\end{equation}

This ordinary differential equation could not be solved
analytically. So we are going to study the numerical solution by
using the observational data from Supernovae Ia, Baryonic
Acoustic Oscillations and the Observational Hubble Data.

\renewcommand{\secname}{Constraints from SNe Ia, BAO and OHD}
\section{Constraints from SNe Ia, BAO and OHD}
\label{sec:\secname}

\subsection{Supernovae Ia}

Supernovae Ia are generally believed to have homogeneous intrinsic
luminosity of peak magnitude. So Supernovae Ia are usually known
as standard candles which could be used to measure the expansion
history of the Universe. The analysis of their distance modulus
versus redshift could provide direct evidence for the acceleration
of the Universe and the analysis also put a constraint on dark
energy models.

Therefore, with the observation of SNe Ia data, we could obtain the
best fit value of the parameter in our model. In order to
constrain the equation of state for the LFDE,
we take the Constitution set with $397$ SNe Ia in terms of the
distance modulus $\mu_{obs}(z)$ compiled in Table $1$ of
Ref.~\cite{Hicken2009}.

For the spatially flat Universe, the Friedmann equation
Eq.~(\ref{eq:fri}) could be written as
\begin{equation}
H(z)=H_{0}\sqrt{\Omega_{m}\left(1+z\right)^{3}+\Omega_{X}}\;,
\label{eq:flatfri2}
\end{equation}
where $\Omega_{m} = \rho_{m}/\rho_{0}$. The luminosity distance
$d_{L}$ of SNe Ia is defined by
\begin{equation}
d_{L}(z)=\frac{c\left(1+z\right)}{H_{0}}F(z)\;.
\end{equation}
The function $F(z)$ is defined by
\begin{equation}
F(z)=\int^{z}_{0}\frac{1}{E(z')}dz'\;,
\end{equation}
with
\begin{equation}
E(z)=\frac{H(z)}{H_{0}}=\sqrt{\Omega_{m}\left(1+z\right)^{3}+\Omega_{X}}\;.
\end{equation}

The theoretical distance modulus is defined by
\begin{equation}
\mu_{th}(z)=m-M=5\log\frac{d_{L}}{10 \textrm{pc}} =
42.39+5\log\frac{1+z}{h}F(z)\;,
\end{equation}
where $m$ and $M$ are the apparent and absolute magnitudes,
respectively. On the other hand, the observational distance
modulus is give by
\begin{equation}
\mu_{obs}=m-M+\alpha(s-1)-\beta c\;,
\end{equation}
where $\alpha$ and $\beta$ are dimensionless parameters, $s$ and
$c$ are shape parameter and color parameter derived from the
fitting to light curves, respectively.

Then we could calculate $\chi^{2}$
\begin{equation}
\chi^{2}_{SNIa}=\sum_{i}^{397}\frac{\left(\mu_{th_{i}}-\mu_{obs_{i}}\right)^{2}}{\sigma^{2}_{i}(\mu_{B})+\sigma^{2}_{int_{i}}}\;,
\end{equation}
where  $\sigma(\mu_{B})$ is the observational variable, which
depends on $\alpha$ and $\beta$. $\sigma(\mu_{B})$ also includes the contribution
of peculiar velocity, $400 km/s$. $\sigma_{int}$ is the intrinsic
dispersion of SNe absolute magnitudes.

\subsection{Baryonic Acoustic Oscillations (BAO)}

The Baryonic Acoustic Oscillations signatures in the large-scale clustering
 of galaxies could act as additional tests for cosmology,
because the acoustic oscillations in the relativistic plasma of
the early Universe could be imprinted onto the late-time power
spectrum of the non-relativistic matter \cite{Eisenstein1998}.
Therefore it could be used to put an additional constraint on dark
energy models.

The BAO relevant distance measure is modelled by volume distance,
which is defined as
\begin{equation}
D_{V}(z)=\left[D_{M}^2(z)z/H(z)\right]^{1/3}\;,
\end{equation}
where $H(z)$ is the Hubble parameter and $D_{M}(z)=\int^{z}_{0}\frac{1}{H(z')}dz'$ is the comoving angular diameter distance.

Eisenstein {\em et al.} \cite{Eisenstein2005} studied the
spectroscopic sample of $46748$ luminous red galaxies from the
Sloan Digital Sky Survey. They found that the combination
$D_{V}(0.35)\sqrt{\Omega_{m}h^{2}}$ has no dependence on the
Hubble constant $H_{0}$ because $D_{V}(0.35)$ is proportional to
$H_{0}^{-1}$. And the combination is well constrained by their
data.

They measured
\begin{equation}
A_{obs}\equiv D_{V}(0.35)\frac{\sqrt{\Omega_{m}H_{0}^{2}}}{0.35c} = 0.469 \pm 0.017(3.6\%)\;.
\end{equation}
For a flat Universe, we have
\begin{equation}
A_{th} = \sqrt{\Omega_{m}}E(z_{1})^{-1/3}\left[\frac{1}{z_{1}}\int^{z}_{0}\frac{1}{E(z')}dz'\right]\;,
\end{equation}
where $z_{1}=0.35$. Therefore $\chi^{2}_{BAO}$ is
\begin{equation}
\chi^{2}_{BAO} = \frac{\left(A_{th}-A_{obs}\right)^{2}}{\sigma^{2}}\;.
\end{equation}

\subsection{The Observational Hubble Data (OHD)}

Relative Galaxy Ages can also be used to constrain cosmological
parameters \cite{Jimenez2002}. Given the measurement of the age
difference of two passively-evolving galaxies formed nearly at the
same time, $\delta$t, and the small redshift interval $\delta$z by
which they are separated, the ratio $\frac{\delta z}{\delta t}$
could be calculated. Then we can infer the derivative:
$\frac{dz}{dt}$.

The quantity measured in the method above is directly related to the Hubble parameter:
\begin{equation}
H(z)_{obs}=-\frac{1}{1+z}\frac{dz}{dt}\;.
\end{equation}

In this paper, we take the observational data in Table V in
Ref.~\cite{Xu2010} from Gemini Deep Survey (GDDS), SPICES and VVDS,
and Keck Observations. We can obtain the theoretical value of
$H(z)$ from the Friedmann equation Eq.~(\ref{eq:flatfri2}).

Therefore $\chi^{2}$ for Hubble data is
\begin{equation}
\chi_{OHD}^{2} = \sum_{i}^{12} \frac{\left(H_{th_{i}}-H_{obs_{i}}\right)^{2}}{\sigma_{i}^{2}}\;.
\end{equation}

Now we have obtained the $\chi^{2}$ of SNIa, BAO and $H_{0}$. Then
the total $\chi^{2}$ is given by:
\begin{equation}
\chi_{tot}^{2}=\chi_{SNeIa}^{2}+\chi_{BAO}^{2}+\chi_{OHD}^{2}\;.
\end{equation}

\renewcommand{\secname}{Results}
\section{Results}
\label{sec:\secname}

\subsection{Best Fit Value of A}

\begin{figure}[here]
\begin{center}
\includegraphics[scale=0.45]{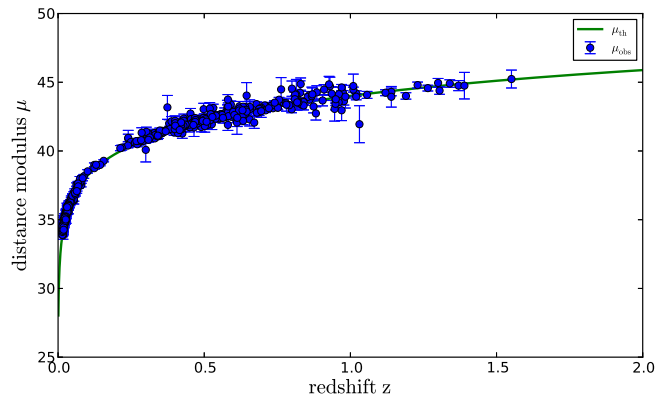}
\end{center}
\caption{The observational distance modulus and the theoretical
distance modulus when $A=-0.3$. The dots with errorbars are for
$\mu_{obs}$ of the $397$ observed SNe Ia data. The solid
line is for the theoretical distance modulus that calculated in a
Universe filled with matter and LFDE.}
\label{fig:mu}
\end{figure}

\begin{figure}[here]
\begin{center}
\includegraphics[scale=0.45]{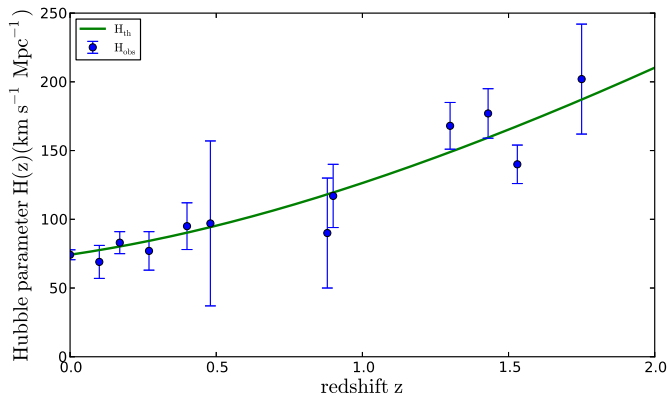}
\end{center}
\caption{The Observational Hubble Data and the theoretical Hubble
parameter when $A=-0.3$. The dots with errorbars are the $12$
observational Hubble data. The solid line is for the theoretical
Hubble parameter that calculated in a Universe filled with dark
matter and LFDE. } \label{fig:Hz}
\end{figure}

Using the joint constraints of SNe Ia, BAO, and OHD, we get the
best fit value of $A:-0.3_{-0.1}^{+0.1}$. In Fig.~\ref{fig:mu}, we
plot the distance modulus for the model and that for the
observational data.
In Fig.~\ref{fig:Hz}, we plot the evolution of the Hubble
parameter with redshifts. It is consistent with the observational
data very well.

\begin{figure}[here]
\begin{center}
\includegraphics[scale=0.45]{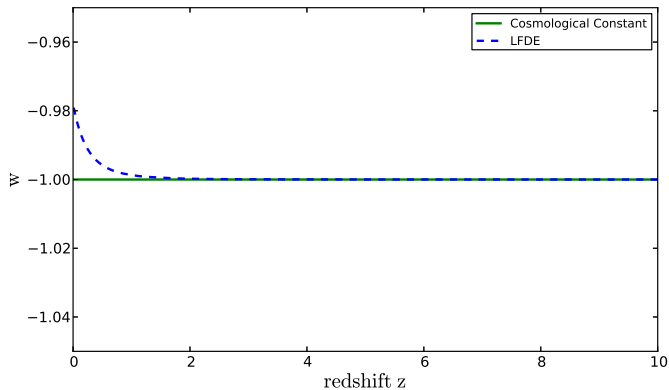}
\end{center}
\caption{The evolution of the equation of state $w
=\frac{p_X}{\rho_X}$. The blue dashed line is the equation of
state for the cosmological constant. The green solid line is the
equation of state for LFDE. When $z>2.0$, $w$ is approaching $-1$.
} \label{fig:w}
\end{figure}

In Fig.~\ref{fig:w}, we plot the equation of state, $w=p_X/\rho_X$
for the lattice fluid.  It is obvious that the equation of state
of LFDE is approaching to $-1$ at the redshifts greater than $2$.
In other words, the lattice fluid behaves as a cosmological
constant at the redshifts greater than $2$. Therefore, we conclude
that the structure formation history would not be modified by this
lattice fluid.

\subsection{Age of the Universe}

The age of the Universe could be calculated from the Friedmann
equation, Eq.~(\ref{eq:fri}). The duration of radiation dominated
epoch is very short compared to the total history of the Universe.
So in order to calculate the age of the Universe, we could simply
consider the two components of the cosmic matter sources: matter
(including dark matter and baryon matter) and dark energy. Using
Eq.~(\ref{eq:flatfri2}) and taking account of the relation
$1+z=\frac{1}{a}$, one find the age of the Universe
\begin{eqnarray}
t_{0}&=&\int_{0}^{t_{0}} dt =
\int_{0}^{\infty}\frac{dz}{H(1+z)}\nonumber\\&=&\int_{0}^{\infty}\frac{dz}{H_{0}(1+z)\sqrt{\Omega_{m}(1+z)^{3}+\Omega_{X}}}\;.
\end{eqnarray}
Then the age of the Universe is found to be $13.5\ \textrm{Gyr}$.

\subsection{The Deceleration Parameter and the Statefinder}

The deceleration parameter $q$ is defined by
\begin{equation}
q\equiv -\frac{\ddot{a}}{aH^{2}}\;.
\end{equation}

By using of the Friedmann equation, Eq.~(\ref{eq:fri}), and the
energy conservation equation
\begin{equation}
\dot{\rho}_{total} = -3H(\rho_{total} + p_{total})\;,
\label{eq:energycon3}
\end{equation}
we can rewrite the deceleration parameter as follows
\cite{Gorini2003}

\begin{equation}
q = \frac{1}{2} +\frac{3}{2} \frac{p_{total}}{\rho_{total}}\;,
\label{eq:qtotal}
\end{equation}
where $p_{total}$ and $\rho_{total}$ are the total pressure and
total energy density of the Universe, respectively.  Since the
total energy density and pressure are contributed by dust
component and LFDE component,
Eq.~(\ref{eq:qtotal}) is actually
\begin{equation}
q = \frac{1}{2} +
\frac{3}{2}\frac{p_{d}+p_{X}}{\rho_{d}+\rho_{X}}=\frac{1}{2} +
\frac{3}{2}\frac{p_{X}}{\rho_{d}+\rho_{X}}\;. \label{eq:qmlf}
\end{equation}
Keeping in mind the equation of state of LFDE, $w=p_X/\rho_X$, and
$\rho_{d}=\rho_{m}(1+z)^3$, we get
\begin{equation}
q = \frac{1}{2} + \frac{3}{2} \frac{w}{1+\frac{\Omega_{m}}{\Omega_{X}}(1+z)^3}\;.
\end{equation}

\begin{figure}[here]
\begin{center}
\includegraphics[scale=0.45]{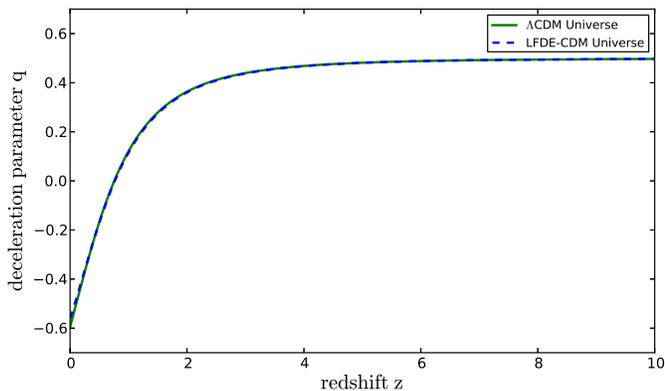}
\end{center}
\caption{The evolution of the deceleration parameter $q$ with the
redshifts. The green solid line is the deceleration parameter of
$\Lambda \textrm{CDM}$ Universe. The blue dashed line is for the
LFDE model. } \label{fig:q}
\end{figure}

We have plotted the evolution of deceleration parameter $q$ with
redshifts in Fig.~\ref{fig:q}. We find it has nearly the same
behavior as the $\Lambda \textrm{CDM}$ model.

The statefinder is introduced by Sahni et al \cite{Sahni2003}. The
definitions are as follows:
\begin{equation}
r\equiv \frac{\dddot{a}}{aH^{3}}\;,\ \ \ s\equiv
\frac{r-1}{3(q-\frac{1}{2})}\;.
\end{equation}
Using the Friedmann equation, Eq.~(\ref{eq:fri}), and the energy
conservation equation, Eq.~(\ref{eq:energycon3}), it is easy to
find
\begin{eqnarray}
r &=& 1-\frac{3\dot{p}}{2\rho\sqrt{\rho}}\sqrt{\frac{3}{8\pi G}}
\;,\nonumber\\ s &=&
-\frac{\dot{p}}{3\rho\sqrt{\rho}}\sqrt{\frac{3}{8\pi G}}\;,
\label{eq:rsfluid}
\end{eqnarray}
where $\rho$ and $p$ stands for the total energy density and total
pressure, respectively. For a two-component matter sources,
Eqs.~(\ref{eq:rsfluid}) take the form
\begin{eqnarray}
r &=& 1+\frac{9}{2(\rho_{1}+\rho_{2})}\left[\frac{\partial
p_{1}}{\partial \rho_{1}}\left(\rho_{1}+p_{1}\right)+\frac{\partial
p_{2}}{\partial \rho_{2}}\left(\rho_{2}+p_{2}\right)\right]\;, \nonumber\\ s &=&
\frac{1}{p_{1}+p_{2}}\left[\frac{\partial p_{1}}{\partial
\rho_{1}}\left(\rho_{1}+p_{1}\right)+\frac{\partial p_{2}}{\partial
\rho_{2}}\left(\rho_{2}+p_{2}\right)\right].
\label{eq:rsfluid2}
\end{eqnarray}

For our case, we have $p_{1}=p_{d}=0$. Then
Eqs.~(\ref{eq:rsfluid2}) become
\begin{eqnarray}
r&=&1+\frac{9}{2} \frac{\rho_{X}+p_{X}}{\rho_{X}+\rho_{d}}
\frac{\partial p_{X}}{\partial \rho_{X}}\;,\nonumber\\
s&=&\frac{\rho_{X}+p_{X}}{p_{X}}\frac{\partial p_{X}}{\partial \rho_{X}}\;.
\label{eq:rsfluid3}
\end{eqnarray}

Finally, from the equation of state for LFDE, Eq.~(\ref{eq:eos}),
we have

\begin{eqnarray}
\frac{\partial p_{X}}{\partial \rho_{X}}=-2\Omega_{X}-
A\left(-\frac{1}{1-\Omega_{X}}+1\right).
\end{eqnarray}

\begin{figure}[here]
\begin{center}
\includegraphics[scale=0.45]{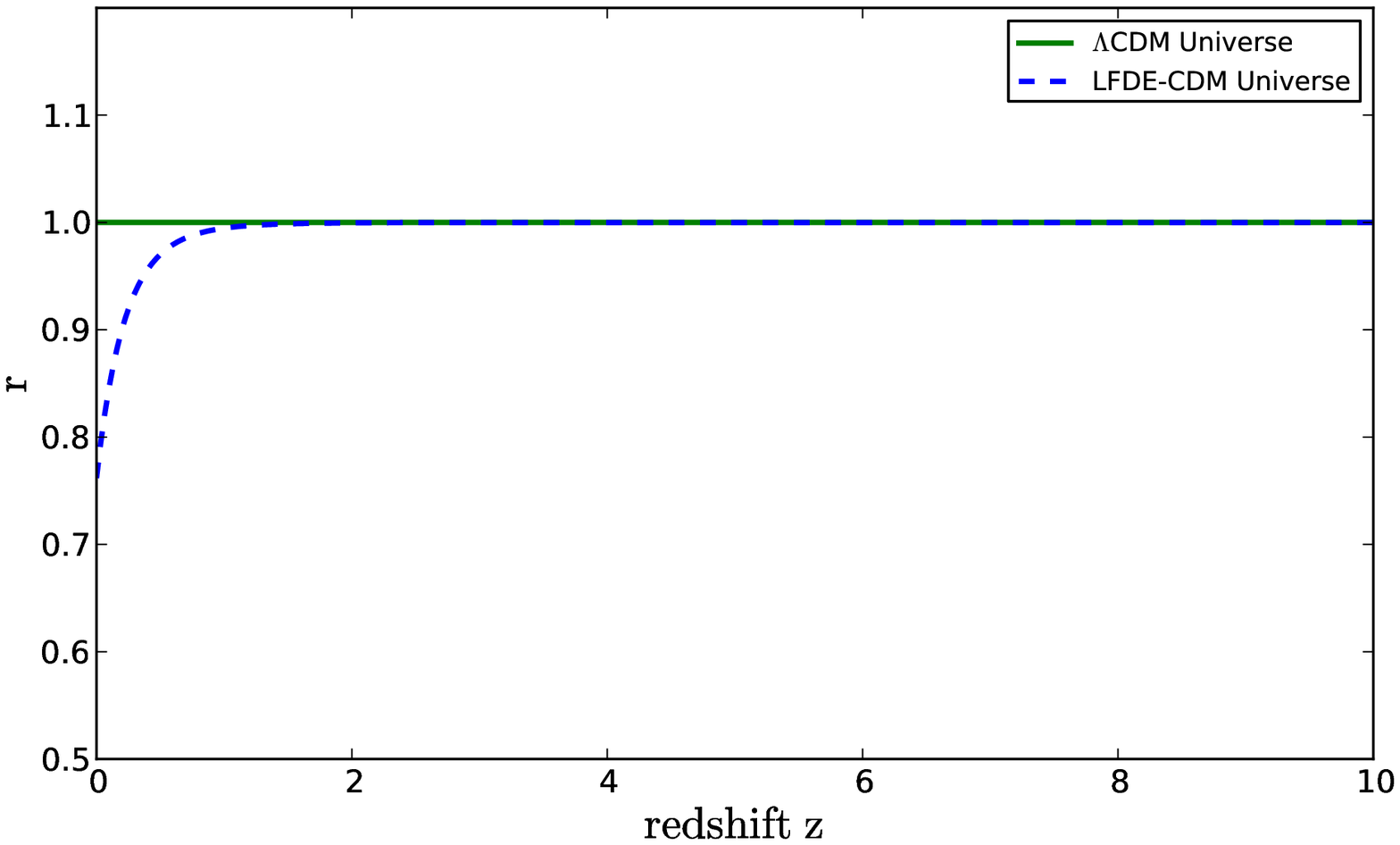}
\end{center}
\caption{The evolution of $r$ with the redshifts. The green solid
line is for the $\Lambda \textrm{CDM}$ model. The blue dashed line
is for the LFDE model. } \label{fig:r}
\end{figure}

\begin{figure}[here]
\begin{center}
\includegraphics[scale=0.45]{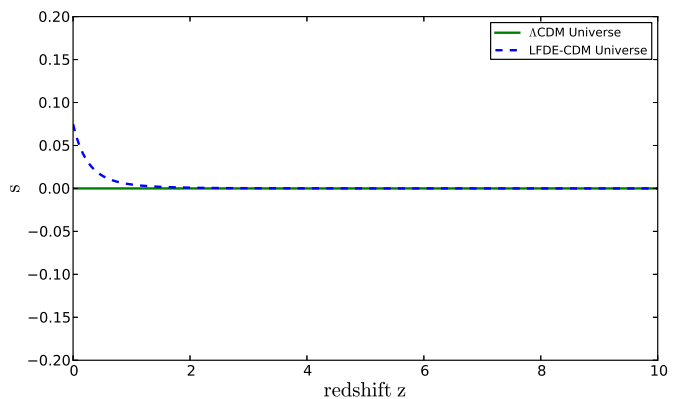}
\end{center}
\caption{The evolution of $s$ with the redshifts. The green solid
line is for the $\Lambda \textrm{CDM}$ model. The blue dashed line
is for the LFDE model.} \label{fig:s}
\end{figure}

The evolution of the statefinder $r$, $s$ are shown in
Fig.~\ref{fig:r} and Fig.~\ref{fig:s}. They show that when the
redshifts are greater than $2.0$, they have almost the same $r$
and $s$ as the cosmological constant. On the other hand, with the
decreasing of redshifts, $r$ and $s$ evolve to the present values:
$r=0.762$, $s=0.075$.

\renewcommand{\secname}{Conclusions}

\section{Conclusions and Discussions}
\label{sec:\secname}

The remarkable discovery of the quantum Hawking radiation
\cite{hawking:1975} reveals that the nature of space, time and
matter is closely related to the statistic thermodynamics. By
turning the logic around, we started from the result of static
thermoplastics in order to seek for the candidate of dark energy.
Sanchez and Lacombe have ever developed a lattice fluid theory
based on a well-defined statistical mechanical model
\cite{Sanchez1978}. One find that the lattice fluid model is so
interesting  that it could perfectly describe the thermodynamic
properties of a wide variety of fluids. Motivated by this point,
we explore the possibility of the lattice fluid as the candidate
of dark energy.

We constrain the model with current cosmological
observational data, including observational data of SNe Ia from the
Constitution set, BAO from SDSS, and Obsevational Hubble Data 
from GDDS, SPICES, VVDS and Keck observations.
 We find the best fit value of parameter $A$:
$A=-0.3_{-0.1}^{+0.1}$. Taking the best value of $A$, we
investigate the comic implications of the model. We find the
equation of state and the statefinder of the lattice fluid are
almost the same as the $\Lambda \textrm{CDM}$ model at the
redshifts greater than $2$. So the structure formation history
would not be modified by this fluid. For the present-day Universe,
we have the equation of state, $w=-0.968$ which is consistent with
many other astronomical observations.

\acknowledgments

Xiaoxian Duan would like to thank Dr. Yan Gong and Prof. Hao Wei
for their help in constraining dark energy models. Thanks also go to
Prof. Xuelei Chen for enlightening advices and Dr. Xin Wang, Bin Yue, Zhigang Li
 for their helpful discussions. This work is
supported by the National Science Foundation of China under the
Key Project Grant 10533010, Grant 10575004, Grant 10973014, and
the 973 Project (No. 2010CB833004).


\begin{thebibliography}{99}




\newcommand\ARNPS[3]{~Ann. Rev. Nucl. Part. Sci.{\bf ~#1}, #2~ (#3)}
\newcommand\AL[3]{~Astron. Lett.{\bf ~#1}, #2~ (#3)}
\newcommand\AP[3]{~Astropart. Phys.{\bf ~#1}, #2~ (#3)}
\newcommand\AJ[3]{~Astron. J.{\bf ~#1}, #2~(#3)}
\newcommand\APJ[3]{~Astrophys. J.{\bf ~#1}, #2~ (#3)}
\newcommand\APJL[3]{~Astrophys. J. Lett. {\bf ~#1}, L#2~(#3)}
\newcommand\APJS[3]{~Astrophys. J. Suppl. Ser.{\bf ~#1}, #2~(#3)}
\newcommand\JHEP[3]{~JHEP.{\bf ~#1}, #2~(#3)}
\newcommand\JCAP[3]{~JCAP. {\bf ~#1}, #2~ (#3)}
\newcommand\LRR[3]{~Living Rev. Relativity. {\bf ~#1}, #2~ (#3)}
\newcommand\MNRAS[3]{~Mon. Not. R. Astron. Soc.{\bf ~#1}, #2~(#3)}
\newcommand\MNRASL[3]{~Mon. Not. R. Astron. Soc.{\bf ~#1}, L#2~(#3)}
\newcommand\NPB[3]{~Nucl. Phys. B{\bf ~#1}, #2~(#3)}
\newcommand\PLA[3]{~Phys. Lett. A{\bf ~#1}, #2~(#3)}
\newcommand\PLB[3]{~Phys. Lett. B{\bf ~#1}, #2~(#3)}
\newcommand\PRL[3]{~Phys. Rev. Lett.{\bf ~#1}, #2~(#3)}
\newcommand\PR[3]{~Phys. Rep.{\bf ~#1}, #2~(#3)}
\newcommand\PRD[3]{~Phys. Rev. D{\bf ~#1}, #2~(#3)}
\newcommand\RMP[3]{~Rev. Mod. Phys.{\bf ~#1}, #2~(#3)}
\newcommand\SJNP[3]{~Sov. J. Nucl. Phys.{\bf ~#1}, #2~(#3)}
\newcommand\ZPC[3]{~Z. Phys. C{\bf ~#1}, #2~(#3)}
 \newcommand\IJGMP[3]{~Int. J. Geom. Meth. Mod. Phys.{\bf ~#1}, #2~(#3)}
  \newcommand\GRG[3]{~Gen. Rel. Grav.{\bf ~#1}, #2~(#3)}
\newcommand\IJMPA[3]{~Int. J. Mod. Phy. A{\bf ~#1}, #2~(#3)}
\newcommand\IJMPD[3]{~Int. J. Mod. Phy. D{\bf ~#1}, #2~(#3)}
\newcommand\CTP[3]{~Commun. Theor. Phys.{\bf ~#1}, #2~(#3)}
\newcommand\CMP[3]{~Comm. Math. Phys.{\bf ~#1}, #2~(#3)}
\newcommand\MPLA[3]{~Mod. Phys. Lett. A.{\bf ~#1}, #2~(#3)}

\bibitem{Riess1998}
A.G.~Riess {\em et al.}, \AJ{116}{1009}{1998}.

\bibitem{Perlmutter1999}
S.~Perlmutter {\em et al.}, \apj {517}, 565 (1999).

\bibitem{Spergel2007}
D.N.~Spergel {\em et al.}, \ Astrophys. J. Suppl. {170}, 377 (2007).

\bibitem{Komatsu2009}
E.~Komatsu {\em et al.}, \ Astrophys. J. Suppl. {180}, 330-376 (2009).

\bibitem{Komatsu2011}
E.~Komatsu {\em et al.}, \ Astrophys. J. Suppl. {192}, 18 (2011).

\bibitem{Eisenstein2005}
D.J.~Eisenstein {\em et al.}, \APJ{633}{560-574}{2005}.

\bibitem{Percival2007}
W.J.~Percival, S.~Cole, D.J.~Eisenstein, R.C.~Nichol, J.A.~Peacock, A.C.~Pope, and A.S.~Stalay, \MNRAS{381}{1053}{2007}.

\bibitem{Percival2009}
W.J.~Percival {\em et al.} , \MNRAS{401}{2148}{2009}.

\bibitem{Jimenez2002}
R.~Jimenez, and A.~Loeb, \APJ{573}{37-42}{2002}.

\bibitem{weinberg:2000}
S. Weinberg, astro-ph/0005265.

\bibitem{Wetterich1988}
C.~Wetterich, \NPB{302}{668}{1988}.

\bibitem{Ratra1988}
B.~Ratra, and J.~Peebles, \PRD{37}{3406}{1988}.

\bibitem{Fujii1982}
Y.~Fujii, \PRD{26}{2580}{1982}; L.H.~Ford, \PRD{35}{2339}{1987}; Y.~Fujii, and T.~Nishioka, \PRD{42}{361}{1990}; T.~Chiba, N.~Sugiyama, and T.~Nakamura, \MNRAS{289}{L5}{1997}.

\bibitem{Carroll1998}
S.M.~Carroll, \PRL{81}{3067}{1998}.

\bibitem{Caldwell2002}
R.R.~Caldwell, \PLB{545}{23}{2002}.

\bibitem{Chiba2000}
T.~Chiba, T.~Okabe, and M.~Yamaguchi, \PRD{62}{023511}{2000}.

\bibitem{Armendariz-Picon2000}
C.~Armendariz-Picon, V.~Mukhanov, and P.J.~Steinhardt, \PRL{85}{4438}{2000}.

\bibitem{Armendariz-Picon2001}
C.~Armendariz-Picon, V.~Mukhanov, and P.J.~Steinhardt, \PRD{63}{103510}{2001}.

\bibitem{Padmanabhan2002}
T.~Padmanabhan, \PRD{66}{021301}{2002}.

\bibitem{Padmanabhan20022}
T.~Padmanabhan, and T.R.~Choudhury, \PRD{66}{081301}{2002}.

\bibitem{Feng05}
B.~Feng, X.L.~Wang, and X.M.~Zhang, \PLB{607}{35}{2005}.

\bibitem{Capozziello2003}
S.~Capozziello, S.~Carloni, and A.~Troisi, \ Recent Res. Dev. Astron. Astrophys. {\bf ~1}, 625 (2003).

\bibitem{Capozziello20022}
S.~Capozziello, \IJMPD{11}{483}{2002}.

\bibitem{Capozziello20032}
S.~Capozziello, V.F.~Cardone, S.~Carloni, and A.~Troisi, \IJMPD{12}{1969}{2003}.

\bibitem{Carroll2004}
S.M.~Carroll, V.~Duvvuri, M.~Trodden, and M.S.~Turner, \PRD{70}{043528}{2004}.

\bibitem{Nojiri2003}
S.~Nojiri, and S.D.~Odintsov, \PRD{68}{123512}{2003}.

\bibitem{Dvali2000}
G.R.~Dvali, G.~Gabadadze, M.~Porrati, \PLB{485}{208}{2000}.

\bibitem{Li04}
M.~Li, \PLB{603}{1}{2004}; 
A.G.~Cohen, D.B.~Kaplan, and A.G.~Nelson, \PRL{82}{4971}{1999}; 
P.~Horava, and D.~Minic, \PRL{85}{1610}{2000}; 
S.~Thomas, \PRL{89}{081301}{2002}; 
S.~Nojiri, and S.D.~Odintsov, \GRG{38}{1285}{2006}; 
X.~Zhang, F.Q.~Wu, \PRD{72}{043524}{2005};
Z.~Chang, F.Q.~Wu, X.~Zhang, \PLB{633}{14}{2006}; 
J.~Zhang, X.~Zhang, and H.~Liu, \PLB{651}{84}{2007}; 
Y.~Ma, X.~Zhang, \PLB{661}{239}{2008}; 
L.~Xu, W.~Li, and J.~Lu, \ arXiv: 0810.4730; 
C.J.~Feng, \PLB{670}{231}{2008}; 
L.N.~Granda, and A.~Oliveros, \PLB{669}{275}{2008}; 
Q.G.~Huang, and M.~Li, \JCAP{0408}{013}{2004}; 
Y.G.~Gong, B.~Wang, and Y.Z.~Zhang \PRD{72}{043510}{2005}; 
B.~Wang, Y.G.~Gong, and E.~Abdalla, \PLB{624}{141}{2005}; 
B.~Chen, M.~Li, and Y.~Wang, \NPB{774}{256}{2007}; 
I.P.~Neupane, \PRD{76}{123006}{2007}; 
J.P.~Wu, D.Z.~Ma, and Y.~Ling, \PLB{663}{152}{2008}; 
H.~Wei, and R.G.~Cai, \ Eur. Phys. J. C {\bf ~59}, 99 (2009); 
Z.~Yi, and T.~Zhang, \MPLA{22}{41}{2007}; 
E.N.~Saridakis, \PLB{660}{138}{2008};
E.N.~Saridakis, \JCAP{0804}{020}{2008}; 
E.N.~Saridakis, \PLB{661}{335}{2008}

\bibitem{Wei07}
H.~Wei, and R.G.~Cai, \PLB{660}{113}{2000}; 
R.G.~Cai, \PLB{657}{228}{2007}; 
Z.~Zhai, T.~Zhang, and W.~Liu, \JCAP{8}{019}{2011}; 
T.~Zhang, C.~Ma and T.~Lan, \ Advances in Astronomy, {2010}, 184284 (2010).

\bibitem{Kamenshchik2001}
A.Y.~Kamenshchik, U.~Moschella, and V.~Pasquier, \PLB{511}{265-268}{2001}.

\bibitem{Bento2002}
M.C.~Bento, O.~Bertolami, and A.A.~Sen, \PRD{66}{043507}{2002}.

\bibitem{Benaoum2002}
H.B.~Benaoum, arXiv: hep-th/0205140v1

\bibitem{Dev2003}
A.~Dev, J.S.~Alcaniz, and D.~Jain, \PRD{67}{023515}{2003}.

\bibitem{Bilic2002}
N.~Bilic, G.B.~Tupper, and R.D.~Viollier, \PLB{535}{17}{2002}.

\bibitem{Alcaniz2003}
J.S.~Alcaniz, D.~Jain, and A.~Dev, \PRD{67}{043514}{2003}.

\bibitem{Chimento2004}
L.P.~Chimento, \PRD{69}{123517}{2004}.

\bibitem{Fabris2011}
J.C.~Fabris, C.~Ogouyandjou, J.~Tossa, and H.E.S.~Velten, \PLB{694}{289-293}{2011}.

\bibitem{Malekjani2011}
M.~Malekjani, A.~Khodam-Mohammadi, N.~Nazari-Pooya, \ Astrophys. Space Sci. {334}, 193-201(2011).

\bibitem{Kremer2003}
G.M.~Kremer, \PRD{68}{123507}{2003}.

\bibitem{Capozziello2002}
S.~Capozziello, S.De~Martino, and M.~Falanga, \PLA{299}{494}{2002}

\bibitem{Capozziello2005}
S.~Capozziello, {\em et al.} \JCAP{04}{005}{2005}.

\bibitem{Sanchez1978}
I.C.~Sanchez and R.H.~Lamcombe, \ Macromolecules {\bf ~11}, 1145 (1978).

\bibitem{bardeen:1973}
J.M.~Bardeen, B.~Carter, and S.W.~Hawking, \CMP{31}{161}{1973}.

\bibitem{hawking:1975}
S.W.~Hawking, \CMP{43}{199}{1975}.

\bibitem{Hicken2009}
M.~Hicken {\em et al.}, \APJ{700}{1097}{2009}.

\bibitem{Eisenstein1998}
D.J.~Eisenstein, and W.~Hu, \APJ{496}{605}{1998}.

\bibitem{Xu2010}
J.~Xu, and Y.~Wang, \JCAP{1006}{002}{2010}.

\bibitem{Gorini2003}
V.~Gorini, A.~Kamenshchik, and U.~Moschella, \PRD{67}{063509}{2003}.

\bibitem{Copeland2006}
E.J.~Copeland, M.~Sami, and S.~Tsujikawa, \IJMPD{15}{1753}{2006}.

\bibitem{Sahni2003}
V.~Sahni, T.D.~Saini and A.A.~Starobinsky, \ U. JETP Lett. {\bf ~77}, 201 (2003).

\end{thebibliography}
\end{document}